\documentclass[aps,pra,twocolumn,superscriptaddress,floatfix,footinbib,notitlepage,groupaddress,showpacs]{revtex4-1} 
\usepackage{times,color,amsfonts,amssymb,amsbsy,amsthm,amsmath,graphics,graphicx,bm,bbm,makecell,mathtools,dsfont,hyperref}
\usepackage{upgreek}
\DeclareFontFamily{OT1}{pzc}{}
\DeclareFontShape{OT1}{pzc}{m}{it}%
              {<-> s * [1.25] pzcmi7t}{}
\DeclareMathAlphabet{\mathpzc}{OT1}{pzc}%
                                 {m}{it}

\let\oldsqrt\sqrt
\def\sqrt{\mathpalette\DHLhksqrt}
\def\DHLhksqrt#1#2{%
\setbox0=\hbox{$#1\oldsqrt{#2\,}$}\dimen0=\ht0
\advance\dimen0-0.2\ht0
\setbox2=\hbox{\vrule height\ht0 depth -\dimen0}%
{\box0\lower0.4pt\box2}}
\begin{document}

\title{Discrete-modulation continuous-variable quantum key distribution enhanced by quantum scissors}

\author{Masoud Ghalaii}
\affiliation{Faculty of Engineering and Physical Sciences, University of Leeds, Leeds LS2 9JT, United Kingdom}
\author{Carlo Ottaviani}
\affiliation{Computer Science and York Centre for Quantum Technologies, University of York, York YO10 5GH, United Kingdom}
\author{Rupesh Kumar}
\affiliation{Department of Physics, University of York, York YO10 5DD, United Kingdom}
\author{Stefano Pirandola}
\affiliation{Computer Science and York Centre for Quantum Technologies, University of York, York YO10 5GH, United Kingdom}
\affiliation{Research Laboratory of Electronics, Massachusetts Institute of Technology (MIT), Cambridge, MA, USA}
\author{Mohsen Razavi}
\affiliation{Faculty of Engineering and Physical Sciences, University of Leeds, Leeds LS2 9JT, United Kingdom}

\begin{abstract}
It is known that quantum scissors, as non-deterministic amplifiers, can enhance the performance of Gaussian-modulated continuous-variable quantum key distribution (CV-QKD) in noisy and long-distance regimes of operation. Here, we extend this result to a {\em non-Gaussian} CV-QKD protocol with discrete modulation. We show that, by using a proper setting, the use of quantum scissors in the receiver of such discrete-modulation CV-QKD protocols would allow us to achieve positive secret key rates at high loss and high excess noise regimes of operation, which would have been otherwise impossible. This also keeps the prospect of running discrete-modulation CV-QKD over CV quantum repeaters alive. 
\end{abstract}

\maketitle

\section{Introduction}
\label{sec:intro}
Quantum key distribution (QKD) is a promising technology for establishing private cryptographic keys between two users \cite{Pirandola:RevQKD2019,Schmitt-Manderbach_Decoy_144km,Yin_MDI_400km}. The security of QKD, which was first introduced in 1984 \cite{bennett2014quantum}, is based on restricting the eavesdropper by the laws of quantum mechanics rather than her ability to efficiently solve certain mathematical problems of high computational complexity \cite{Gisin_QCrypRev_2002}. If properly implemented, this makes QKD secure against the most powerful computers now and in the future.

QKD can be implemented using a number of optical techniques, the most well-known genre of which relies on encoding the key  bits on, e.g., the polarization of single photons, among other discrete degrees of freedom of optical signals.
Continuous-variable QKD (CV-QKD) protocols, such as the Gaussian-modulated technique proposed by Grosshans and Grangier in 2002 (GG02) \cite{Grosshans_GG02_PRL,Grosshans_GG02_Nature}, are introduced as an alternative class, where coherent communication techniques, such as homodyne or heterodyne detection, are employed \cite{Hirano_PulsedHOM,Yonezawa_QTele_Exp,Yokoyama_NatPhoton}. In a CV-QKD protocol, data is encoded on the quadratures of an optical field \cite{Grosshans_GG02_PRL,Grosshans_GG02_Nature,Braunstein_QICVRev_2005,Cerf_Leuchs_Polzik,Weedbrook_GaussQIRev_2012}. 

The progress in implementing CV-QKD protocols has been noteworthy in the past few years \cite{Diamanti_Rev15,Diamanti_Rev16}. 
This has been facilitated by removing some of the security challenges arisen from regenerating the local oscillator \cite{Qi_LLOS,Huang_LLOS,Soh_LLOS} at the receiver, and by the involvement of some commercial actors \cite{Laudenbach_Huawei} to further deploy such technologies. 
Despite this progress, it is generally believed that CV-QKD is perhaps a good option for short-distance or low-loss links \cite{Pirandola2015}, while discrete-variable QKD could be more suitable for long distances. This is partly because of the difficulties with implementing highly efficient reconciliation algorithms for CV-QKD, as well as the less developed quantum repeater paradigms for CV systems.

The scope for long-distance CV-QKD has, however, changed with some recent developments in the field. For instance, one solution is to use non-deterministic amplification \cite{Blandino_CVQKD_idealNLA,Zhang_NLA_EB_CVQKD,Xu_4S_NLA_CVQKD,Ghalaii_CVQKD_QS}. It has been shown that by using a realistic implementation of an amplification device, e.g., a quantum scissor (QS) \cite{Ghalaii_CVQKD_QS, Pegg_QSs,Ralph_Lund_QSNLA}, the security distance of Gaussian-modulated CV-QKD protocols can be increased. {Quantum scissors have already been demonstrated experimentally \cite{Ferreyrol_ImpNLA_PRL, Barbieri_NLA_Exp_Rev} and used for entanglement distillation \cite{Xiang_NatPhys}.} Using quantum scissors, or similar ideas, the first generation of CV quantum repeaters have then been proposed \cite{Dias_Ralph_CVQRs, Furrer_Munro_CVQRs, Guha_CV_Repeater}. Another technique that can potentially improve the rate-versus-distance behavior in CV-QKD protocols is to use a non-Gaussian discrete modulation  \cite{Leverrier_DMCVQKD_PRL,Becir_DMCVQKD_8CS,Papanastasiou_QKDwithPCCS,Lin_dmCVQKD,Leverrier_DMCVQKD_PRA}. It is generally perceived that, especially, at low signal-to-noise ratio levels, which we have to deal with at long distances, it would be easier to design an error correction scheme for discrete-modulation encoding as opposed to the Gaussian one \cite{Leverrier_DMCVQKD_PRA,Leverrier_SNR}. 

In this paper, we consider all above enabling factors within a single setup to study the rate-versus-distance behavior for a discrete-modulation CV-QKD system that uses quantum scissors at its receiver. This is effectively the main building block in the quantum repeater setup proposed in Ref.~\cite{Dias_Ralph_CVQRs}, which, in our work, is used for discrete-modulation CV-QKD. A realistic analysis of our setup could then be used to assess the practicality of the proposed repeater setups. It has already been shown that, by using an ideal non-deterministic linear amplifier (NLA) at the receiver’s side, one can increase the maximum transmission distance and tolerable excess noise of the quadrature-phase-shift-keying (QPSK) protocol \cite{Xu_4S_NLA_CVQKD}. However, a study that accounts for a realistic NLA, such as a quantum scissor, is missing. This is important, because one of the key incentives for using discrete-modulation CV-QKD is its similarity with existing coherent optical communications systems, which possibly makes its adoption and implementation more straightforward. It is also important to consider a physical realization of the NLA in our system, as opposed to measurement-based ones \cite{Fiurasek_VirNLA,Walk_CVQKD_postsel,Chrzanowski_MBNLA}, because otherwise the system cannot be used in a repeater setup. Measurement-based NLAs often offer lower key rates when used in CV-QKD setups \cite{Zhao_VirNLA}, which is another reason for considering the physical deployment of a QS in our setup. For further clarification on this matter, interested readers are referred to the discussions in Ref.~\cite{Ghalaii_CVQKD_QS}.

The security analysis of discrete-modulation CV-QKD has turned out to be more challenging than its Gaussian counterpart. The reported analysis in Ref.~\cite{Leverrier_DMCVQKD_PRL} relies on the linearity of the channel for its security. But, the authors admit that this is not an easy condition to verify. In order to rectify this problem, in Ref.~\cite{Leverrier_DMCVQKD_PRA}, they come up with a modified scheme in which they can relax the assumption on the channel linearity by requiring Alice to send three types of signals: Gaussian modulated ones for channel estimation, discrete-modulation ones for key generation, and a range of decoy states to conceal the discrepancy between the latter two in the eyes of an eavesdropper. The decoy states would, effectively, make the modulated signals look Gaussian, which makes the security analysis more manageable. This approach, however, to a large extent, takes away the practical aspects of discrete-modulation CV-QKD. Very recently, new analyses have emerged, which rely on numerical optimization of the key rate based on certain constraints obtained from the measurement results \cite{PRXGhorai, Norbert_DM_CV}. In our setup, we have another complication that results from using the QS, which is non-deterministic. This would further make the channel non-Gaussian, which implies that the optimal attack by an eavesdropper could also be non-Gaussian. By carefully engineering our system to remain close to Gaussian, we can, however, obtain a reasonable estimation of the secret key rate by restricting the eavesdropper to Gaussian attacks {enabled by an entangling cloner \cite{Navascues_EntanglingCloner}}. This allows us to use a thermal-loss model for the channel, for which we calculate the key rate. We show how the performance of our non-Gaussian CV-QKD system is enhanced in this case, especially in high-loss and high-excess noise regimes. 

The outline of the paper is as follows. 
In Sec.~\ref{sec:sys_desc}, we describe the system under study.
In Sec.~\ref{sec:secret_key_analysis}, we present the key rate analysis of the QS-assisted CV-QKD protocol with non-Gaussian modulation. 
We then discuss our numerical results in Sec.~\ref{sec:numerical_results} and conclude our paper in Sec.~\ref{sec:conclusion}.  

\section{System description} 
\label{sec:sys_desc}
In this section, we present our proposed QS-amplified CV-QKD protocol with  discrete modulation and its equivalent entanglement-based (EB) version. Both schemes are depicted in Fig.~\ref{fig:setup}. Different components of the system are described below.

\subsection{Modulation and Detection}
In a conventional non-Gaussian/discrete modulation protocol, a particular finite constellation of coherent states is considered and used for encoding data. A constellation of four and eight coherent states are the well-known cases \cite{Leverrier_DMCVQKD_PRL,Leverrier_DMCVQKD_PRA,Xu_4S_NLA_CVQKD,Becir_DMCVQKD_8CS,Papanastasiou_QKDwithPCCS}. In this study, we focus on the QPSK protocol. 
We assume that the sender, Alice (A), sends her prepared signals to the receiver, Bob (B), via a quantum channel. In our proposed protocol, however, Bob is equipped with a single QS in order to amplify the received signal. Bob applies the QS operation just before his homodyne detection, which are both owned and handled by him. 
The homodyne measurement results are recorded whenever the QS operation is successful. 

\begin{figure}[t]
\centering
\includegraphics[width=.9\linewidth]{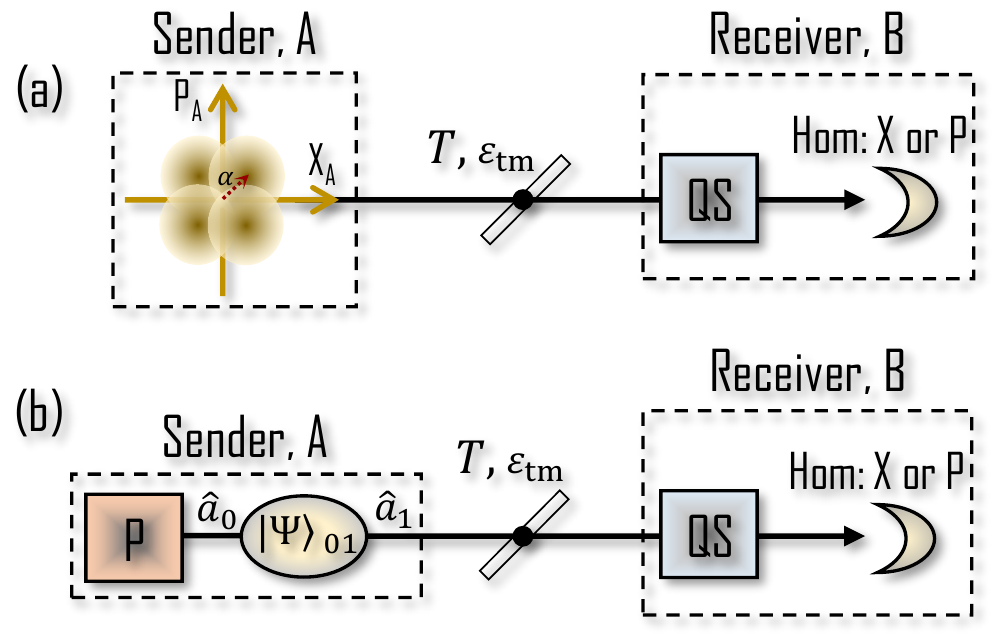} 
\caption{System description. (a) Schematic view of discrete-modulation CV-QKD protocol equipped with a quantum scissor as part of its receiver. Here, the four yellow circles at the sender side represent the constellation of the four coherent states used at the encoder. (b) The entanglement-based CV-QKD protocol equivalent to (a). The quantum channel is modeled by the equivalent excess noise at the transmitter side, represented by $\varepsilon_{\rm tm}$, and its transmissivity $T$. $|\Psi \rangle_{01} $, QS, Hom and P boxes, respectively, represent the bipartite entangled state in Eq.~\eqref{EB-state}, a probabilistic quantum scissor as seen in Fig.~\ref{fig:single-qs}, the homodyne detection and projective measurement modules in $\{|\psi_k\rangle_0\}$ basis.}
\label{fig:setup}
\end{figure}

More precisely, the prepare and measure (P\&M) version of the protocol runs as follows. First, Alice randomly chooses a coherent state from the set $\{|\alpha_k\rangle=|\alpha e^{(2k+1)i\pi/4} \rangle \}_{k=0}^{3}$, with $\alpha \in \mathbbm{R}^+$, and sends it to Bob through a quantum channel; see Fig.~\ref{fig:setup}(a). Such a constellation can be generated by rotation of a coherent state in the position-momentum phase space. 
The parameter $\alpha $ can be optimized to give the maximum secret key rate.
In addition, we assume $\alpha_k=(x_{Ak}+ip_{Ak})/2, k=0,\dots,3$, with real parameters $x_{Ak}$ and $p_{Ak}$ being chosen randomly according to the following uniform probability mass functions: $f_{X_A}(x_{Ak})=f_{P_A}(p_{Ak})=1/4$.
At the receiver, Bob randomly measures one quadrature, $\widehat{x}_B=\widehat{a}_B^\dagger+\widehat{a}_B$ or $\widehat{p}_B=i(\widehat{a}_B^\dagger-\widehat{a}_B)$, of the QS output using homodyne detection, where $\widehat{a}_B^\dagger$ represents the creation operator for the output mode of the QS.
The trusted parties, Alice and Bob, keep the detection results only if the QS operation is successful in the respective round; that is, only one of detectors D1 or D2, in Fig.~\ref{fig:single-qs}, clicks. 
By doing reconciliation and privacy amplification, the parties can then obtain a common string of secret bits.

In order to calculate the secret key generation rate, especially the Holevo information term, it is often easier to consider the equivalent EB scheme, which is shown in Fig.~\ref{fig:setup}(b). In the EB version, instead of randomly choosing and sending single-mode coherent states, Alice measures one mode of a bipartite entangled state, and sends the other one to Bob. In the Gaussian modulation case, the employed entangled state is a two-mode squeezed vacuum (TMSV) state, and Alice measurement is heterodyne detection. In the case of the QPSK protocol, it has been shown that one can start with a TMSV state, and apply a certain measurement to obtain the following state \cite{Leverrier_DMCVQKD_PRA}
\begin{align}
\label{EB-state}
|\Psi \rangle_{01} =& \sum_{k=0}^{3} \sqrt{\lambda_k} |\phi_k\rangle_0 |\phi_k\rangle_1 \nonumber \\
= & \frac{1}{2}  \sum_{k=0}^{3} |\psi_k\rangle_0 |\alpha_k\rangle_1 ,
\end{align}
where
\begin{align}
|\phi_k\rangle = \frac{-\frac{\alpha^2}{2}}{\sqrt{\lambda_k}} \sum_{n=0}^{\infty} (-1)^n \frac{\alpha^{4n+k}}{\sqrt{(4n+k)!}} |4n+k \rangle  \nonumber
\end{align}
and
\begin{align}
|\psi_k\rangle_0 = \frac{1}{2} \sum_{m=0}^{3} e^{(2k+1)im\pi/4} |\phi_m \rangle_0   \nonumber
\end{align}
are orthogonal non-Gaussian states, with $\lambda_{0,2}=e^{-\alpha^2/2}\big( \cosh(\alpha^2) \pm \cos(\alpha^2) \big)/2$ and $\lambda_{1,3}=e^{-\alpha^2/2}\big( \sinh(\alpha^2) \pm \sin(\alpha^2) \big)/2$. The subscripts $0$ and $1$ refer to optical modes represented by $\widehat a_0$ and $\widehat a_1$, respectively. In the procedure described in Ref.~\cite{Leverrier_DMCVQKD_PRA}, there is a chance that instead of the state in Eq.~\eqref{EB-state}, we end up with a decoy state. In this paper, we focus only on the key generation part, which results from the state in Eq.~\eqref{EB-state}, and do not consider the parameter estimation task, for which we should either send Gaussian modulated states \cite{Leverrier_DMCVQKD_PRA}, or use numerical techniques \cite{PRXGhorai}. In the end, the equivalence of P\&M and EB schemes of the protocols is obtained via a proper projective measurement $\widehat{P}$ in $\{|\psi_k\rangle_0\}$, $k=0,\ldots,3$, basis.

\subsection{Quantum Channel}
\label{sec:channel_desc}
The parties are assumed to use a thermal-loss channel with transmittivity $T$ and an excess noise $\varepsilon$.
A potential model for such a channel is given by a beam splitter, with transmissivity $T$, that mixes Alice's signals and the eavesdropper's thermal state, given by the following expression:
\begin{align}
\label{Eq:thermal}
\widehat{\rho}_{\mathsf{th}} =  \int d^2\beta \frac{e^{-\frac{|\beta |^2}{\varepsilon/2}}}{\pi \varepsilon/2} |\beta \rangle_{\widehat{a}_{\rm N}} \langle \beta |,
\end{align}
where $\widehat a_N$ is the annihilation operator corresponding to the noise port, and $d^2\beta = d \Re{\beta} d \Im{\beta}$. The equivalent excess noise at the input to the channel is then given by $\varepsilon_{\rm tm} = (1-T)\varepsilon /T$. 

In principle, the parties cannot tell what kind of channel they have without proper parameter estimation. {As we will explain in Sec.~\ref{sec:secret_key_analysis},} the assumption of a thermal-loss channel corresponds to the case of a Gaussian attack {enabled by an entangling cloner}, which may not be optimal for our non-Gaussian system. 
However, as long as the system does not deviate considerably from the Gaussian framework, the results obtained are expected to provide us with a reasonable estimate of the potential key rate \cite{Malaney_Sat} that can be obtained by a more rigorous analysis. We use the above model to calculate the relevant parameters of the co-variance matrix when QSs are in use. 

\subsection{Quantum Scissors}

\begin{figure}[t]
\centering
\includegraphics[width=.55\linewidth]{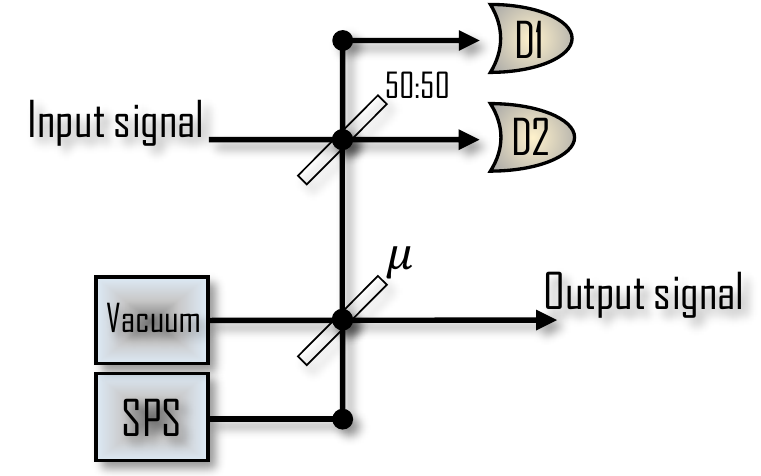}
\caption{The schematic view of a quantum scissor. Here, we assume
that a ready-to-shoot ideal single-photon source (SPS) is in use,
and that the single-photon detectors have unity efficiencies. The QS amplification gain is defined as $g=\sqrt{(1-\mu)/\mu}$.}
\label{fig:single-qs}
\end{figure}

Quantum scissors are at the core of the NLA module proposed by Ralph and Lund \cite{Ralph_Lund_QSNLA}. A single QS has two beam splitters in its setup, one of which is balanced while the other has a transmittance $\mu$; see Fig.~\ref{fig:single-qs}. The 50:50 beam splitter couples the incoming signal to a single photon that has gone through the imbalanced beam splitter. A click on exactly one of detectors D1 and D2 would herald success of the QS. We note that an on-demand ideal single photon source assumed here in our analysis. 

Here we obtain the output state of the QS, upon successful operation, for an input state $\widehat{\rho} = \frac{1}{4} \sum_{k=0}^{3} |\alpha_k\rangle \langle \alpha_k | $ to the thermal-loss channel described in Sec.~\ref{sec:channel_desc}. In order to do so, we use the results reported in Ref.~\cite{Ghalaii_CVQKD_QS}, in which the output state of such a setup for an arbitrary coherent state at the input has been derived. We then obtain 
\begin{align}
\label{QS-state}
\widehat{\rho}_{\mathsf{QS}} (\alpha)= &  a(\alpha) |0\rangle_1 \langle 0| + c(\alpha) |1\rangle_1 \langle 1| ,
\end{align}
where $\widehat{\rho}_{\mathsf{QS}} (\alpha)$ is the density matrix at the output of the QS upon successful operation and
\begin{align}
\label{state_coeff}
\begin{cases}
a(\alpha)= \frac{2\mu [2F(2F+1)+T|\alpha|^2]}{ (2F+1)^3 P^{\mathsf{PS}} (\alpha)} e^{-\frac{T|\alpha|^2}{2F+1} }   \\
c(\alpha)=   \frac{2(1-\mu)}{ P^{\mathsf{PS}} (\alpha)} \Big( \frac{e^{-\frac{T|\alpha|^2}{2F+1}}}{2F+1}  - \frac{e^{-\frac{T|\alpha|^2}{2F}}}{4F}   \Big) ,
\end{cases}
\end{align}
with $F=\frac{1}{2}+ \frac{1}{4} T\varepsilon_{\mathsf{tm}}$. In Eq.~\eqref{state_coeff},  
\begin{align}
\label{postsel_prob}
P^{\mathsf{PS}}(\alpha) = &\frac{2[(2F+1)^2 -\mu (2F+1)+\mu T|\alpha|^2] }{(2F+1)^3} e^{-\frac{T|\alpha|^2}{2F+1}} \nonumber \\
& - \frac{1-\mu}{2F} e^{-\frac{T|\alpha|^2}{2F}} \nonumber \\
= & P_{\rm succ}(\alpha)/2, 
\end{align}
where $P_{\rm succ}(\alpha)$ is the success probability for the QS. 

An interesting observation from Eq.~\eqref{QS-state} is that the output state of the QS is non-Gaussian. This is not just because we have used non-Gaussian modulation, but even for a single coherent state at the input, as discussed in Ref.~\cite{Ghalaii_CVQKD_QS}, the output state is in the subspace spanned by $\{|0\rangle, |1\rangle \}$. There are two implications for this behavior. First, the QS amplification cannot be noise free, as in an ideal NLA, but the amount of noise can vary based on the input signal and the amplification gain. Further, this non-Gaussianity can complicate the security analysis of the protocol.  In our work, we manage this additional complexity by restricting the eavesdropper (Eve) to collective Gaussian attacks \cite{Pirandola:ColGausAttacks_PRL2008}, as we will discuss in Sec.~\ref{sec:secret_key_analysis}.

The non-Gaussianity of the channel manifests itself in the statistics that we can obtain from Bob's homodyne measurement. In particular, using similar techniques as in Ref.~\cite{Ghalaii_CVQKD_QS}, the output probability distribution of $\widehat{x}_B$-quadrature can be calculated as follows:
\begin{align}
\label{output_dist}
f_{X_B}(x_B) &= \mathrm{tr} [ \widehat{\rho}_{\mathsf{QS}} (\alpha)  |x_B \rangle \langle x_B | ] \nonumber \\
&=  \big[ a(\alpha) + 2 c(\alpha) x_B^2 \big]  \frac{e^{-x_B^2} }{\sqrt{\pi}}, 
\end{align}
with $\widehat{x}_B |x_B\rangle = x_B |x_B\rangle$. As can be seen in Eq.~\eqref{output_dist}, similar to the Gaussian-modulation case, the output probability distribution function is composed of a Gaussian and a non-Gaussian term. In the regime, where $a(\alpha) \gg c(\alpha)$, we are very close to a fully Gaussian system. For this to happen $\alpha$ needs to be small. In the other extreme, when $c(\alpha) \gg a(\alpha)$, we get a bimodal form for the output distribution, which is clearly non-Gaussian. {A similar observation, although via a different technique, has been made in earlier experiments on QSs, where the asymmetry in the measured Wigner functions grows with increase in the intensity of the input state \cite{Ferreyrol_ImpNLA_PRL}.}

Similarly, we can work out the conditional output probability distribution:
\begin{align}
\label{output_dist-cond}
f_{X_B}(x_B|x_{Ak})= \mathrm{tr} [  \widehat{\rho}_{\mathsf{QS},c} (x_{Ak}) |x_B \rangle \langle x_B | ] ,
\end{align}
where  
\begin{align}
\widehat{\rho}_{\mathsf{QS},c} (x_{Ak})= & a_c(x_{Ak}) |0\rangle_1 \langle 0| +  b_c (x_{Ak}) |0\rangle_1 \langle 1| \nonumber \\
& + b_c^\ast (x_{Ak}) |1\rangle_1 \langle 0|  + c_c(x_{Ak})  |1\rangle_1\langle 1|  
\end{align}
{is the QS output state conditioned on Alice sending a signal with $X$ quadrature $x_{Ak}$ {\em and} observing a click on D1.} In this case,
\begin{align}
\begin{cases}
a_c(x_{Ak})= \frac{2\mu \big( 4F(2F+1)+T(\alpha^2+2x_k^2)\big) }{(2F+1)^3 P^{\mathsf{PS}}_c(x_{Ak})} e^{-\frac{T(\alpha^2+2x_k^2)}{2(2F+1)}} \\
b_c(x_{Ak})= - \frac{2 \sqrt{\mu(1-\mu)T}x_k}{(2F+1)^2 P^{\mathsf{PS}}_c(x_{Ak}) }e^{-\frac{T(\alpha^2+2x_k^2)}{2(2F+1)}} \\
c_c(x_{Ak})=1-a_c(x_{Ak})  
\end{cases} 
\end{align}
and
\begin{align}
P^{\mathsf{PS}}_c(x_{Ak})= &\frac{2 (2F+1)^2- 2\mu(2F+1) + \mu T (\alpha^2+2x_k^2) }{(2F+1)^3}  \nonumber \\
& \times e^{-\frac{T(\alpha^2+2x_k^2)}{2(2F+1)}} - \frac{1-\mu }{2F} e^{-\frac{T(\alpha^2+2x_k^2)}{4F}}.
\end{align}
We will later use the above expressions in order to calculate the mutual information between the parties.

\section{Secret Key Rate Analysis}
\label{sec:secret_key_analysis}
In this section, we present the key rate analysis for our QS-equipped QKD system. We calculate the secret key generation rate for our system under the assumption that the eavesdropper is limited to Gaussian attacks. {That is, we assume that the eavesdropper replaces the channel with an entangling cloner, where one part of a TMSV state is coupled, at a beam splitter, with Alice's signal and sent to Bob, while the other part would be retained by Eve and will be measured once Alice and Bob have sifted their data.} In this case, we can assume that the effective channel between Alice and Bob is a thermal-loss channel as we described in Sec.~\ref{sec:channel_desc}. {Note that, the key rate obtained in this case is not necessarily a lower bound on the key rate in the most general case because the optimal attack by an eavesdropper can be non-Gaussian. That is, for a given joint state between Alice and Bob, the required purification by Eve may not be obtained by an entangling cloner. Assuming that Eve uses an entangling cloner, however, at each run of the protocol, the state between Alice, Eve, and Bob, before the QS, is pure. Now because in the QS operation we make a projective measurement, the conditional state between Alice, Eve, and Bob, after the QS, is also pure. This is exactly the same state by which we calculate the Holevo information component of the key rate.} As it is pointed out in Refs.~\cite{Malaney_Sat}, the key rate obtained in our case is expected to be a close approximation to a true lower bound on the key rate for the nominal joint state obtained by Alice and Bob. 

In the asymptotic limit of many runs of the protocol, the secret key rate of a CV-QKD protocol under collective attack is given by \cite{Cerf_Leuchs_Polzik}
\begin{align}
\label{rate}
K= \beta I_{AB} - \chi_{EB} ,  
\end{align}
where $\beta$, $I_{AB}$, and $\chi_{EB}$ are, respectively, the reconciliation efficiency, the mutual information between the parties, and the leaked/accessible information to Eve when reverse reconciliation is used. However, since the QS is a non-deterministic operation, the key rate should be multiplied by the average probability of success, $P_{\mathsf{succ}}(\alpha)$, where all possible inputs are considered in the averaging. Therefore, the secret key rate reads as follows
\begin{align}
\label{rate-QS}
K_{\mathsf{QS}} \geq P_{\mathsf{succ}}(\alpha) (\beta I_{AB} - \chi_{EB}).
\end{align}
In our protocol, we discard data associated to the unsuccessful events and use only the post-selected data in order to produce a secret string of bits. 
In the following, we first derive the exact value for $I_{AB}$, in Sec.~\ref{subsec:mut_info}, and an upper bound for $ \chi_{EB}$, in Sec.~\ref{subsec:Hol_info}, for the thermal-loss channel. 

\subsection{Mutual Information}
\label{subsec:mut_info}
By definition, the mutual information of two random variables $X_A$ and $X_B$ is the difference between the entropy function $H(X_B)$ and the conditional entropy $H(X_B|X_A)$:
\begin{align}
\label{mutinf_def}
I_{AB}= H(X_B)-H(X_B|X_A),
\end{align}
where
\begin{align}
\label{Ent-XB}
H(X_B)= \int dx_B ~ f_{X_B}(x_{B}) \log_2 \frac{1}{f_{X_B}(x_{B})}
\end{align}
and
\begin{align}
\label{Ent-XB-cond}
H(X_B|X_A) =  \frac{1}{4} \sum_{k=0}^{3} \int dx_B ~ f_{X_B}(x_B|x_{Ak}) \log_2 \frac{1}{f_{X_B}(x_B|x_{Ak})}.
\end{align}
Functions $f_{X_B}(x_{B})$ and $f_{X_B}(x_B|x_{Ak})$ are given in Eqs.~(\ref{output_dist}) and (\ref{output_dist-cond}), using which and the above equations, we numerically calculate the mutual information. We note that the input quadrature is a discrete random variable whereas the output is, in principle, continuous.

\subsection{Holevo Information} 
\label{subsec:Hol_info}
\begin{figure}[t]
	\centering
	\includegraphics[width=.9\linewidth]{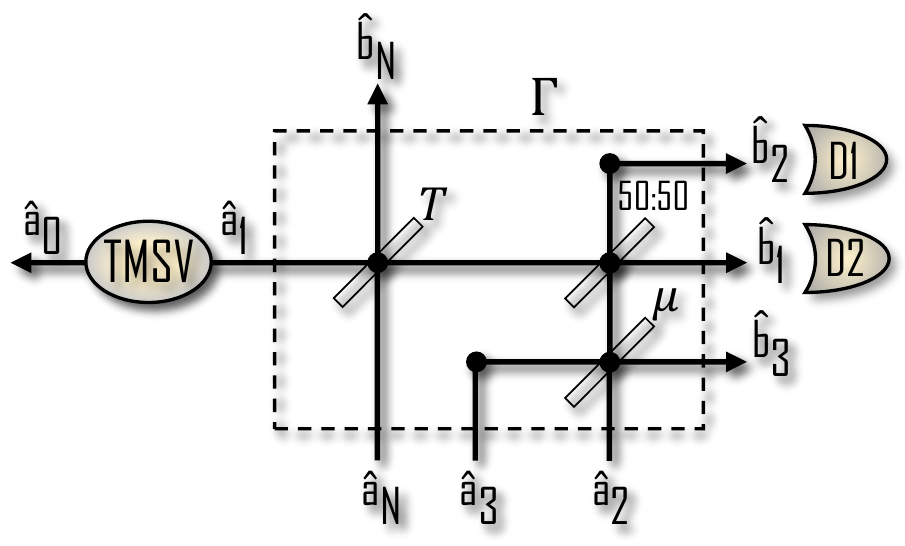}
	\caption{Entanglement-based version of the QS-amplified CV-QKD scheme. The noisy quantum channel and the QS are considered as a combined system, with input modes $\widehat{a}_1-\widehat{a}_3$, and $\widehat{a}_{\rm N}$, and output modes $\widehat{b}_1-\widehat{b}_3$, and $\widehat{b}_{\rm N}$. The initial state of modes represented by $\widehat{a}_0-\widehat{a}_1$ is given by $|\Psi \rangle_{01}$. The initial state of the modes represented by operators $\widehat{a}_2$, $\widehat{a}_3$, and $\widehat{a}_{\rm N}$ is, respectively, given by a single photon, a vacuum, and the thermal state in Eq.~\eqref{Eq:thermal}. } 
	\label{fig:EB-CV-QKD} 
\end{figure}

We upper bound the leaked information, $ \chi_{EB}$, by calculating the Holevo term for a Gaussian channel with the same co-variance matrix (CM) {between Alice and Bob's quadratures} as that of our system \cite{Garcia-Patron_OptGaussAttacks,Navascues_OptGaussAttacks}. 
In order to find the CM, in the case of our thermal-loss channel, we first need to find the bipartite state between Alice mode $\widehat{a}_0$ and Bob mode $\widehat{b}_3$ for the proposed QPSK setup in Fig.~\ref{fig:EB-CV-QKD}. In doing so, we let mode $\widehat{a}_1$ of the state in Eq.~\eqref{EB-state} to propagate through the noisy quantum channel, which we model via a beam splitter, with transmissivity $T$, which couples Alice's signal to the thermal state in Eq.~\eqref{Eq:thermal},  and subsequently undergoes the QS operation. 
Quantum scissors involve a measurement as they are successful if only one of their detectors clicks. We define measurement operator $\widehat{M}=(\mathbbm{1} -|0 \rangle_1 \langle 0|) \otimes|0\rangle_2 \langle 0|$, corresponding to a click on detector D1 and no click on D2, where $\mathbbm{1}$ represents the identity operator for optical mode entering D1, and $|0 \rangle_1$ and $|0 \rangle_2$ are vacuum states corresponding to, respectively, optical modes $\widehat{b}_1$ and $\widehat{b}_2$.

In order to calculate the joint state of modes $\widehat{a}_0$ and $\widehat{b}_3$, we follow the same procedure as in Ref.~\cite{Ghalaii_CVQKD_QS} that relies on finding input-output characteristic functions for the module $\Gamma$ in Fig.~\ref{fig:EB-CV-QKD}. Upon a successful QS operation, i.e., $\widehat{M}$ measurement, we obtain
\begin{align}
\label{bipartite-state}
\widehat{\rho}_{03}= \frac{1}{4 P^{\mathsf{PS}}} \sum_{k=0}^{3} \sum_{l=0}^{3} |\psi_k\rangle_0\langle \psi_l | \otimes \widehat{\Omega}_3^{kl} ,
\end{align}
where
\begin{align}
\label{state_b3}
\widehat{\Omega}_3^{kl}= \int \frac{d^2\xi_3}{\pi} \zeta_A^{kl}(\xi_3) \widehat{D}_N(\widehat{b}_3,\xi_3)
\end{align}
is the state that Bob measures, with $\widehat{D}_N(\widehat{b},\xi)= e^{\xi\widehat{b}^{\dagger}} e^{-\xi^\ast\widehat{b}}$ being the normally-ordered displacement operator of mode $\widehat{b}$. In Eq.~\eqref{state_b3}, 
\begin{align}
\zeta_A^{kl}(\xi_3)=\int  \frac{d^2\xi_1}{\pi} \frac{d^2\xi_2}{\pi} \chi_A^{kl}(\xi_1,\xi_2,\xi_3) 
\end{align}
where, for $|\alpha_k\rangle_1 \langle \alpha_l|$ as the input state,
\begin{align}
\label{anti-norm-func}
\chi_A^{kl}(\xi_1,\xi_2,\xi_3) = & 
e^{-F|\xi_1-\xi_2|^2} e^{\sqrt{\frac{T}{2}} [ \alpha_l^\ast (\xi_1 - \xi_2) - \alpha_k (\xi_1^\ast - \xi_2^\ast) ]}   \nonumber \\
&\times e^{-\frac{\mu}{2}   |\xi_1 + \xi_2 + \sqrt{2} {g} \xi_3|^2 }  e^{-\frac{1-\mu}{2}   |\xi_1 + \xi_2 - \sqrt{2}/g \xi_3|^2 }  \nonumber \\
& \times  (\pi \delta^2(\xi_1)-1)  \big( 1-\frac{\mu}{2}   |\xi_1 + \xi_2 + \sqrt{2} {g} \xi_3|^2 \big)
\end{align}
is the antinormally-ordered characteristic function of the output states in Fig.~\ref{fig:EB-CV-QKD} after tracing over the noise mode $\widehat{b}_{\rm N}$, which belongs to a potential eavesdropper. Also, success probability for measurement $\widehat{M}$ is given by 
\begin{align}
P^{\mathsf{PS}}  = & \frac{1}{4} \sum_{k=0}^{3} \int \frac{d^2\xi_1}{\pi} \frac{d^2\xi_2}{\pi}  \chi_A^{kk}(\xi_1,\xi_2,0) \nonumber \\
= & \frac{1}{4} \sum_{k=0}^{3} \zeta_A^{kk}(0) = \zeta_A^{00}(0), 
\end{align} 
where $\zeta_A^{kl}(0)$ is given by Eq.~\eqref{H_func}. This result exactly matches that of the P\&M scheme, given in Eq.~\eqref{postsel_prob}. We remark that the total success probability is given by $P_{\mathsf{succ}}=2 P^{\mathsf{PS}} =2 \zeta_A^{00}(0)$, which also accounts for the case of D2 clicking and D1 not clicking.

Next, in order to find a lower bound on the secret key rate, following original works in \cite{Leverrier_DMCVQKD_PRL,Leverrier_DMCVQKD_PRA}, we use the optimality of Gaussian collective attacks in the asymptotic limit for a given CM \cite{Garcia-Patron_OptGaussAttacks,Navascues_OptGaussAttacks}.  
Now that the bipartite state between Alice and Bob is given by Eq.~\eqref{bipartite-state}, we can work out the first and second order moments in the CM, which is turned out to be in the standard symplectic form \cite{Weedbrook_GaussQIRev_2012} below:
\begin{align}
\label{CM-TMSV}
V_{AB}=
\left(\begin{array}{cc}
V_x \mathbbm{1} & V_{xy} \sigma_{\mathsf{z}} \\
V_{xy} \sigma_{\mathsf{z}} &  V_y \mathbbm{1}
\end{array}\right),
\end{align}
where $\mathbbm{1}=\text{diag}(1,1)$ and $\sigma_{\mathsf{z}}=\text{diag}(1,-1)$ are Pauli matrices. 
In Appendix~\ref{app:CM-parameters}, we derive the closed form expression of the triplet $(V_x,V_{xy},V_y)$. Note that the obtained CM, in the case of having a successful QS operation for vacuum state at the input, i.e., when $\alpha=0$,  results in identity CM, i.e., $V_{AB}=\mathbbm{1}\otimes \mathbbm{1}$, as one would expect.   
Having found the CM, one can then work out a bound on Holevo information using the set of equations given in Appendix~\ref{app:HolevoInfo}.

\begin{figure}[t]
	\centering
	\includegraphics[scale=.55]{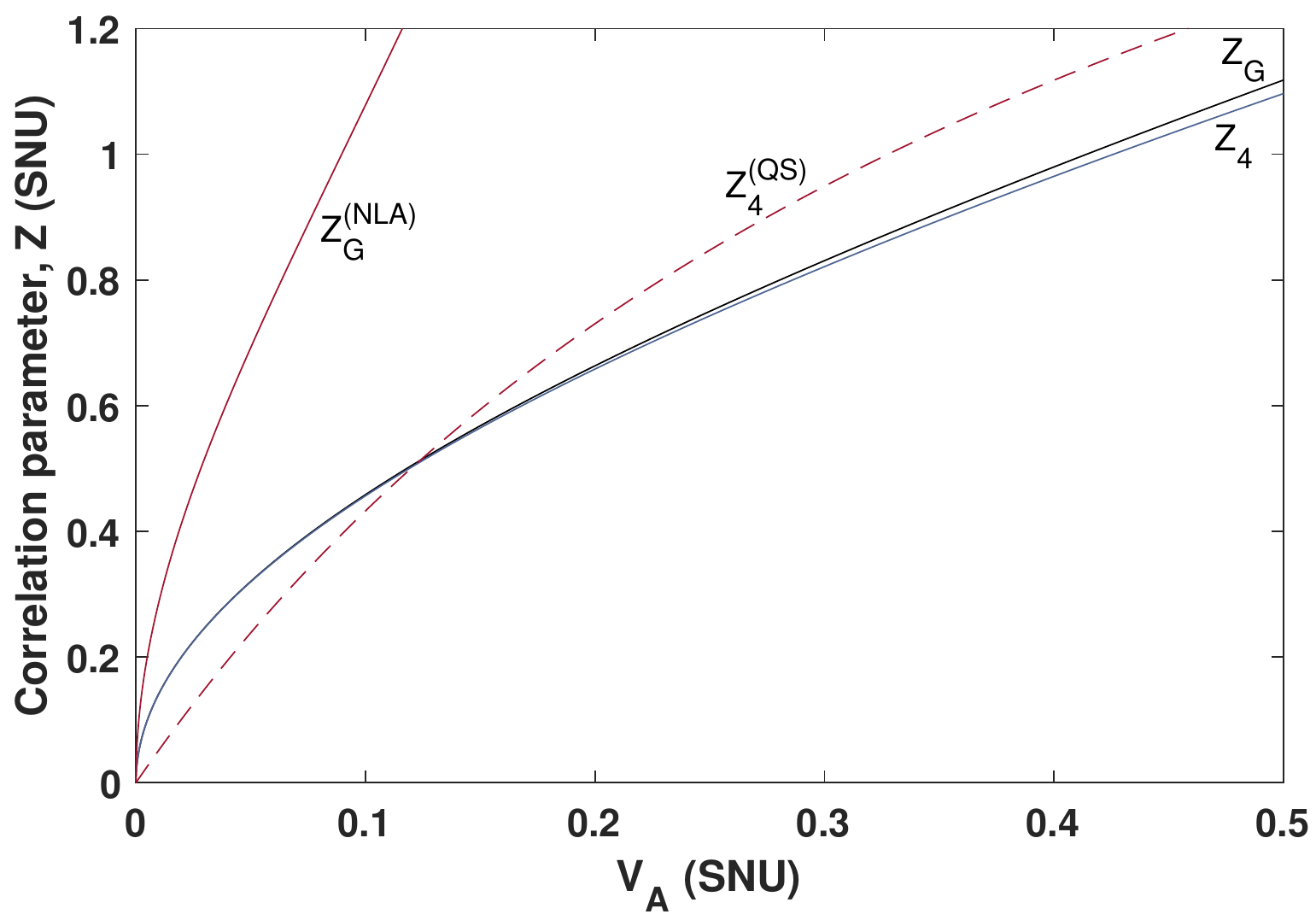}
	\caption{Correlation factor for the GG02 protocol (solid black), the four coherent-state constellation without (solid blue) and with (dashed red) a QS with amplification gain $g=2$. The solid red curve belongs to the TMSV state amplified via an ideal NLA ($g=2$); see text for more information. Here, the channel is assumed loss-less and without any excess noise.}
	\label{fig:Z_correlation}
\end{figure}

An important feature of the CM in Eq.~\eqref{CM-TMSV} is its correlation parameter, defined as $Z_{4}^{\rm (QS)}=V_{xy}/\sqrt{T}$, which characterizes the amount of correlation between the parties's quadratures upon a successful QS operation. 
Figure~\ref{fig:Z_correlation} compares $Z_{4}^{\rm (QS)}$ in our QS-based system with that of the no-QS setup, $Z_4$, in \cite{Leverrier_DMCVQKD_PRA}, and then compares both with that of the Gaussian modulation case without ($Z_{\rm G}$) and with ($Z_{\rm G}^{\rm (NLA)}$) an ideal NLA. In the case of Gaussian modulation without an NLA, instead of $|\Psi\rangle_{01}$, we start with a TMSV state given by $\sqrt{1-\lambda^2} \sum_{n=0}^{\infty} \lambda^n |n\rangle_0 |n\rangle_1$, for which the corresponding CM is given by $
\left(\begin{array}{cc}
(V_A+1) \mathbbm{1} & Z_{\rm G} \sigma_{\mathsf{z}} \\
Z_{\rm G} \sigma_{\mathsf{z}} &  (V_A+1) \mathbbm{1}
\end{array}\right)$, with
$Z_{\rm G}=\sqrt{V_A^2+2V_A}$, where $V_A=2 \lambda^2 /(1-\lambda^2)$ is its corresponding modulation variance. The parameter $\lambda$ in the above TMSV state would ideally change to $g \lambda$ once one arm of the TMSV state goes through an ideal NLA with gain $g$ \cite{Ralph_Lund_QSNLA}. The corresponding correlation term, $Z_{\rm G}^{\rm (NLA)}$, can then be calculated by $\sqrt{(V'_A)^2+2V'_A}$, where $V'_A=2 g^2\lambda^2 /(1-g^2\lambda^2)$. 

Figure \ref{fig:Z_correlation} compares the above four correlation parameters as a function of $V_A$. In the case of the QPSK protocol, $V_A=2\alpha^2$.  We can see that $Z_{4}^{\rm (QS)}$ overtakes the two no-NLA curves at a $V_A$ around 0.15. This suggests that the amount of correlation between the trusted parties' signals has been enhanced by the use of a QS. This may imply that higher key generation rates can be obtained in certain regimes of operation. One should, however, note that by increasing $V_A$, hence $\alpha$, we may reduce the success probability of the QS system. Furthermore, by increasing $\alpha$, Eve's Gaussian attack would be further away from her optimal attack. We will discuss this point in our numerical results when we optimize the secret key rate over system parameters. One final interesting point in Fig.~\ref{fig:Z_correlation} is that the correlation term for the ideal NLA is always better than the QS system. This may suggest that the earlier analysis that rely on an ideal NLA may overestimate what can be achieved with a realistic NLA system.

\section{Numerical Results}
\label{sec:numerical_results}
In this section, we present some numerical results for the secret key rate of our QS-amplified QPSK CV-QKD system and compare it with that of the no-QS protocol, and its Gaussian modulated (GM) variants. To that end, we solve a dual optimization problem. 
We find the maximum value for the lower bound in Eq.~\eqref{rate-QS} by optimizing over $\alpha$, which specifies the modulation variance, and the QS parameter $g$, which specifies the QS amplification gain. 
In our numerical results, for a channel with length $L$, we assume that $T=10^{-\kappa L /10}$, where $\kappa= 0.2$ dB/km is the loss factor for optical fibers. Also, we nominally assume a reconciliation efficiency equal to one and that Bob, upon successful QS events, uses an ideal homodyne detection, with no electronic noise, to measure the received signals.

\begin{figure}[t]
	\includegraphics[scale=.55]{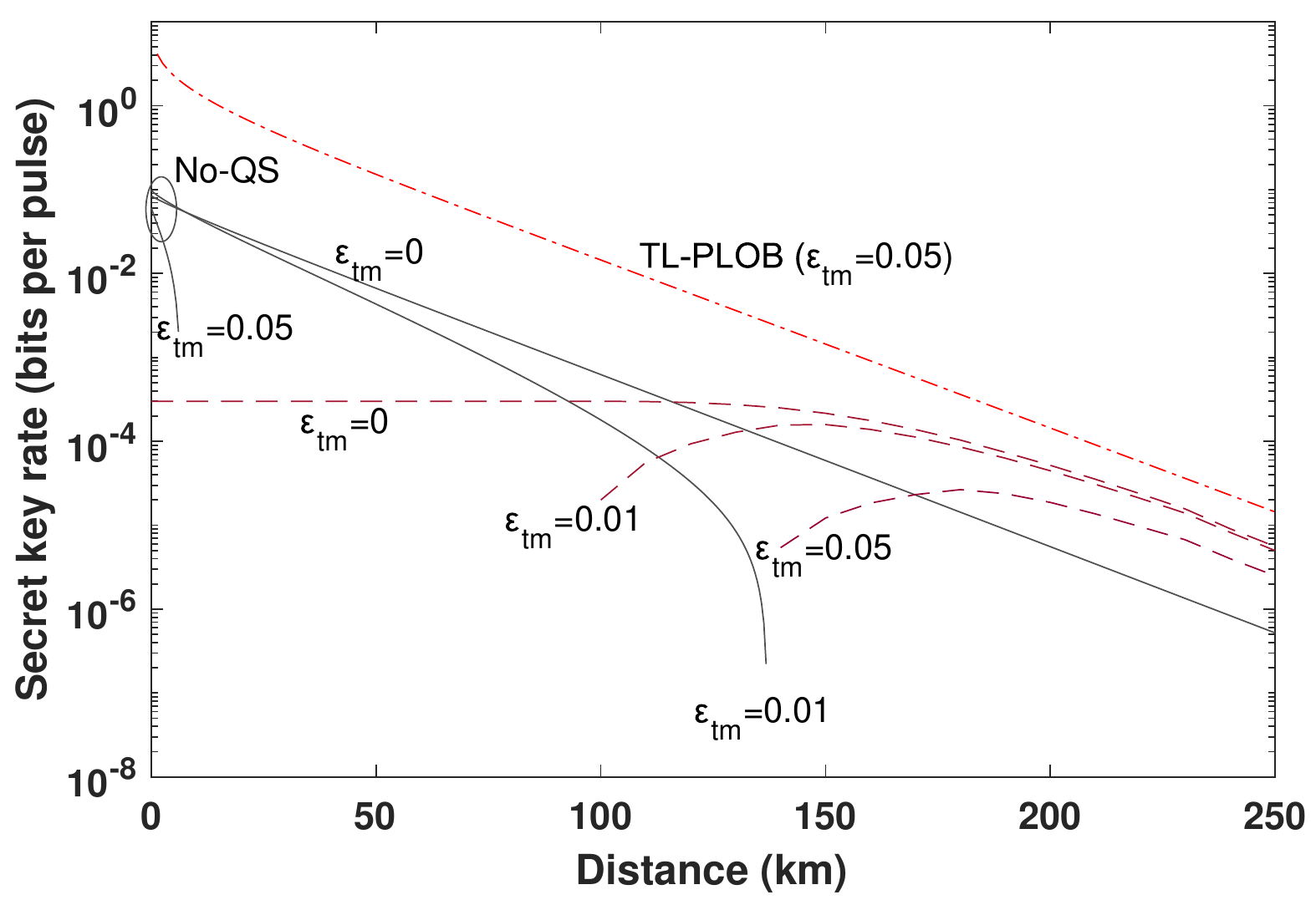}
	\caption{Numerical results of the optimized secret key rate for QS-equipped QPSK modulation CV-QKD protocol versus distance (dashed lines), as compared to that of the protocol with no-QS (solid lines). The ultimate thermal-loss PLOB bound \cite{Pirandola_PLOB17} is shown at the top.}
	\label{fig:rates}
\end{figure}

Figure~\ref{fig:rates} shows the optimized key rates for the no-QS \cite{Leverrier_DMCVQKD_PRL,Leverrier_DMCVQKD_PRA} and QS-equipped discrete modulation protocols versus distance. We observe that the behavior of the different curves shown in Fig.~\ref{fig:rates} is very much akin to the Gaussian modulation QS-equipped CV-QKD presented in Ref.~\cite{Ghalaii_CVQKD_QS}. In particular, the QS-based systems are capable of beating their no-QS counterparts after a certain distance, and considerably increase the maximum security distance achievable by the underlying QKD protocol. The crossover distance at an input excess noise equal to 0 and 0.01 shot-noise units (SNU) is, respectively, around 120~km and 110~km. In the case of $\varepsilon_{\rm tm} = 0.05$, the no-QS system has a very low reach, whereas, by using a QS, the system can now provide positive secret key rates at distances over 140~km. It can also be seen that the QS based system offers either zero or very low secret key rates at short distances. This, as pointed out in Ref.~\cite{Ghalaii_CVQKD_QS}, can be because of the additional noise by the QS, especially, for large inputs, which requires us to use much lower values of $\alpha$ that would be used in the no-QS system. This could make the signal component, at short distances, less than the excess noise part, hence resulting in no secure keys.

The opposite effect is seen at long distances where QS-based systems are offering a key rate parallel to the fundamental bounds for secret key generation rate for a thermal-loss channel (labeled by TL-PLOB). This is the bound given in Eq.~(23) of Ref.~\cite{Pirandola_PLOB17} at an equivalent mean thermal photon number, $\bar n =  \varepsilon_{\rm tm} T/(2(1-T))$, to our receiver excess noise (here at $\varepsilon_{\rm tm} = 0.05$) \cite{Pirandola_IOP2018}. This extended security distance suggests that once the input to the QS is low enough, which is at long distances, the post-selection offered by the QS can improve the signal-to-noise ratio to a level that positive secret key rates are distillable. We have numerically verified that positive key rates are indeed achievable for $\varepsilon_{\rm tm} < 0.09$ for the QS-based system.

\begin{figure}[t]
	\includegraphics[scale=.5]{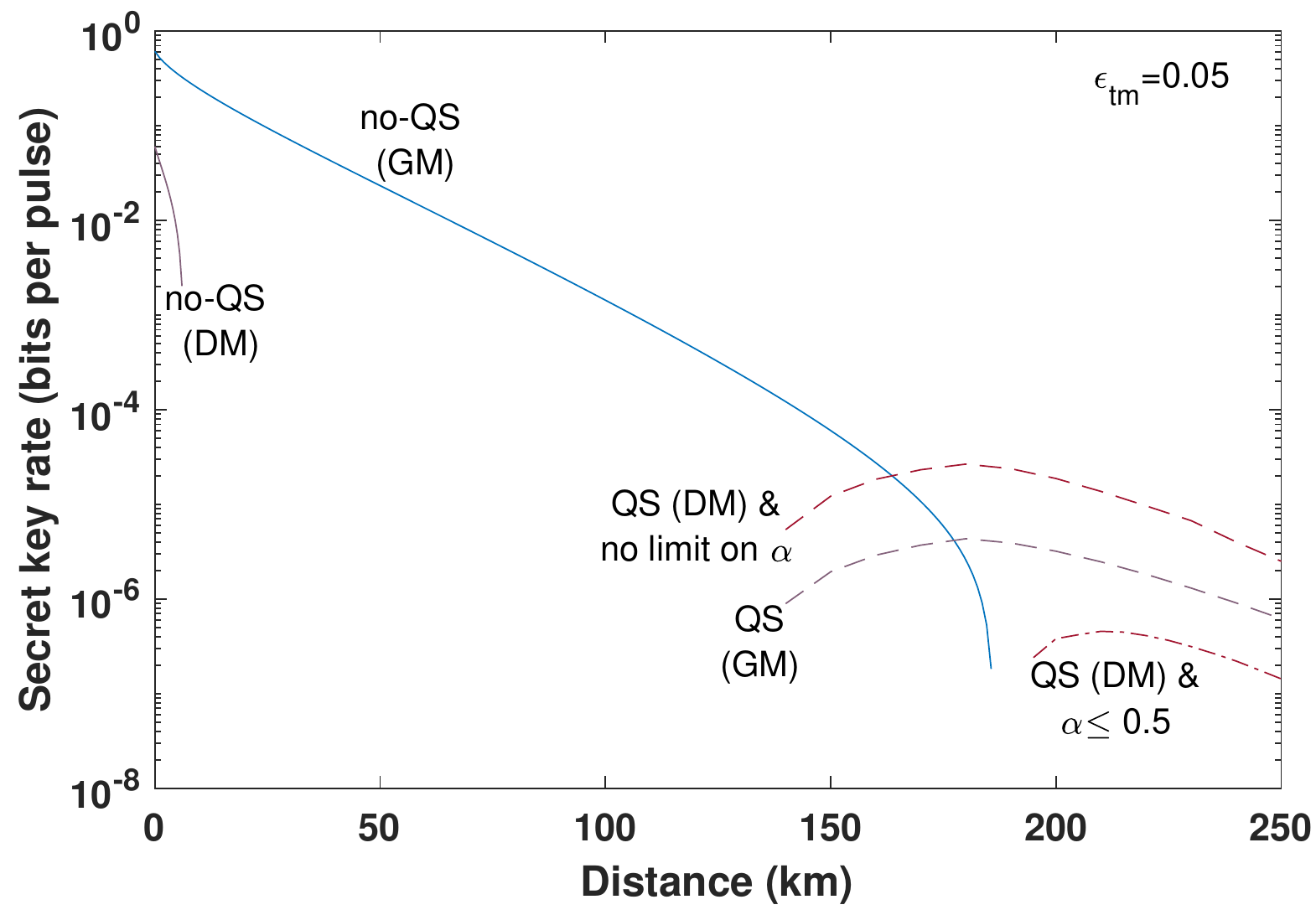}
	\caption{Numerical results of the optimized secret key rate for discrete modulation (DM) CV-QKD protocol versus distance, as compared to that of the Gaussian modulated (GM) GG02 protocol with and without a QS. The lower curve represents the result of optimized key rate when $\alpha$ is capped at 0.5. The rates are obtained at $\beta=1$.}
	\label{DM-vs-GM}
\end{figure}

\begin{figure}[t]
	\centering
	\includegraphics[scale=.5]{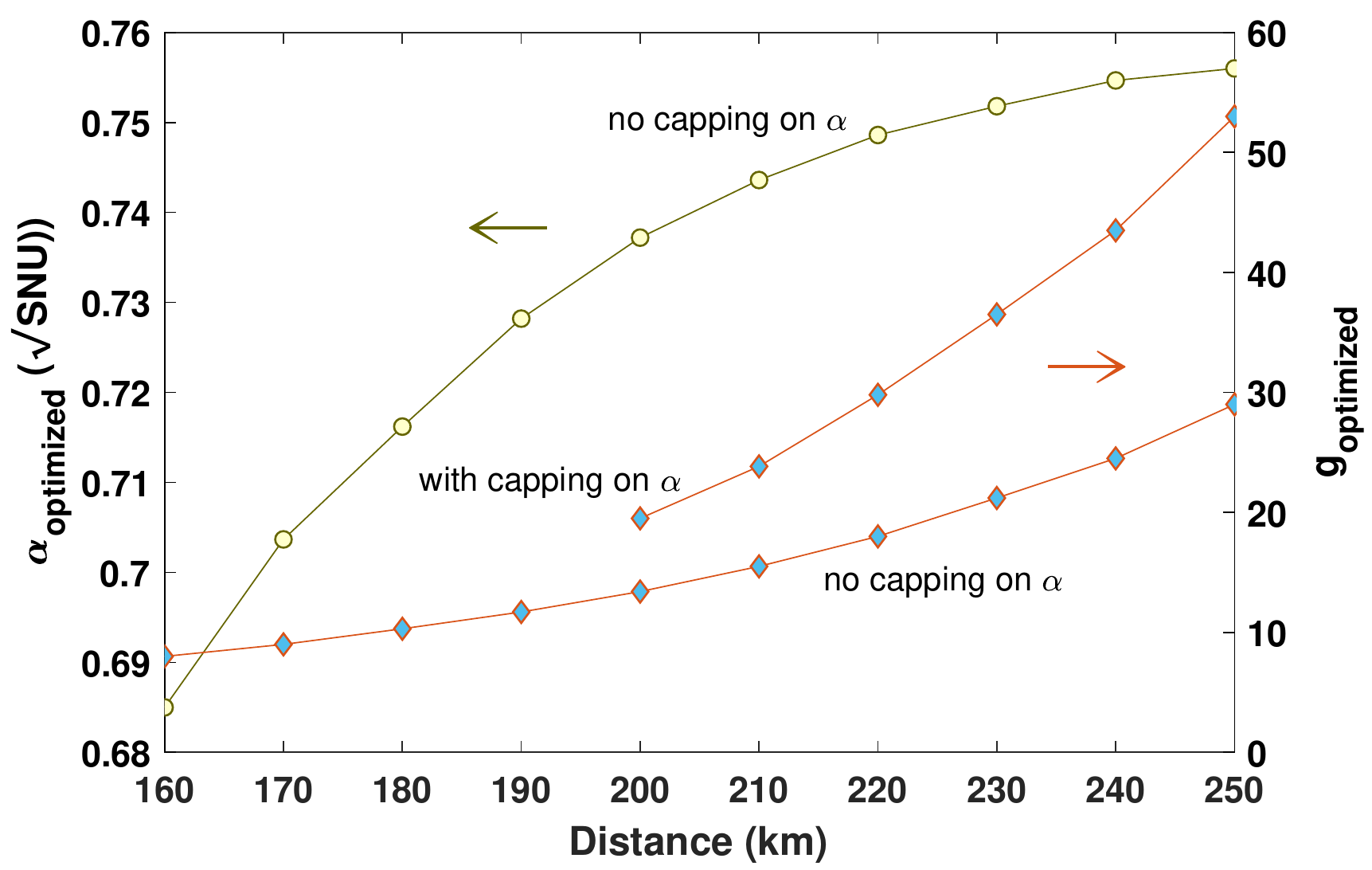}
	\caption{Optimized input amplitude (marked by circles) and optimized amplification gain (marked by diamonds) versus channel length at $\varepsilon_{\rm tm} = 0.05$ with an without a cap (0.5, not shown on the graph) on $\alpha$.}
	\label{fig:optimization}
\end{figure}

The QS-equipped discrete modulation (DM) system in this work seems to offer more resilience to excess noise and channel loss than its GM counterpart considered in Ref.~\cite{Ghalaii_CVQKD_QS}. For instance, the maximum tolerable excess noise in the latter case is around 0.06 SNU as compared to 0.09 SNU in the former case. The secret key rate obtained at a high excess noise value of 0.05 SNU is also higher for the DM versus GM case. This has been shown in Fig.~\ref{DM-vs-GM} where the secret key rate for both systems, in the presence and absence of a QS, has been shown. This result is, however, counter-intuitive, and must be taken with caution. There is a fundamental difference between the GM and DM case in that the latter is not a Gaussian modulation especially for large values of $\alpha$. As shown in Fig.~\ref{fig:optimization}, the optimal value of $\alpha$ is around 0.7 at $\varepsilon_{\rm tm} = 0.05$. In our analysis, we have, however, assumed that Eve is restricted to a Gaussian attack, which will become less optimal as the input modulation deviates further from a Gaussian one. What our numerical results would then suggest is that {for an Eve restricted to an entangling cloner,} it is better to use a non-Gaussian modulation as this would make Eve's attack even less optimal. 

If we want to obtain a more realistic account of what a non-restricted Eve could achieve in our system, we should then cap the choice of $\alpha$ in our optimization to a value that preserves the Gaussianity of the input signal to some good extent. A suggested cap for $\alpha$ is given in \cite{PRXGhorai} to be around 0.5. The lower curve in Fig.~\ref{DM-vs-GM} shows the secret key rate under this constraint, while the corresponding optimal value of $g$ is shown in Fig.~\ref{fig:optimization}. It is now clear that the rate obtained for the DM case, at $\beta = 1$, is lower than that of the GM case. The no-QS GM system will, however, offer no positive key rate for $\beta < 0.98$, which implies that, if one considers the more efficient reconciliation techniques for DM systems, there would be regimes of operation where the DM system outperforms the GM case.  Note that, as shown in Fig.~\ref{fig:optimization}, by capping $\alpha$, larger values of gain is needed by the QS to achieve the optimal key rate.

Finally, we would like to comment on the suitability of quantum scissors in CV quantum repeaters. One of the objectives of calculating the key rate of a QS equipped CV-QKD system was the similarity of the setup to what was proposed, as the main building block, in recent proposals for CV repeaters \cite{Dias_Ralph_CVQRs, Guha_CV_Repeater}. Our intuition was that if a realistic QS could not offer any advantage over the no-QS one, then the prospect of a CV repeater that relies on such QS devices would also be questionable. Our results suggest that there are regimes of operation that QS-based systems offer some advantage. We are, however, short of a convincing argument that such regimes of operation would be those in which repeater systems could operate as well. In fact, while our results keep the prospect of functioning CV repeaters open, they also highlight the importance of considering all noise effects before jumping into any conclusions. Our analysis could then be used to further study the proposed repeater setups and assess how, in practice, they can perform.

\section{Conclusions and Discussion}
\label{sec:conclusion}
In this work, we studied the performance of a CV-QKD system that used quadrature phase shift keying modulation at the encoder and a certain optical state truncation device, i.e., a quantum scissor, before its homodyne receiver. The objective was to find if and to what extent the use of a QS, as a non-deterministic amplifier, could improve the rate behavior of the system at long distances. We showed that, by optimizing the relevant system parameters, the QS-equipped system could tolerate more excess noise than the no-QS discrete-modulation system, and therefore could reach longer distances at positive values of excess noise. This effect was similar to that of a Gaussian-modulated CV-QKD system \cite{Ghalaii_CVQKD_QS}, but in the discrete-modulation case we observed additional tolerance against excess noise if only Gaussian attacks are considered, or assume lower reconciliation efficiencies for the Gaussian modulation case, as is often the case in practice. This enables us to extend the reach of CV-QKD systems provided that we supplement them with additional devices such as single-photon sources and single-photon detectors \cite{Senellart_qdots:NatPhot2017,Cahall_SPDetectors:2017}. This, at first, may sound counterproductive as it takes away some of the practical advantages of CV-QKD systems. But, one should note that these additional equipment are only needed at the receiver end of the link, which, in a practical setup, can represent a shared network node in a quantum network. Moreover, our analysis would specify the range of distances for which the use of a quantum scissor could be beneficial. Over shorter distances, one could still use a conventional system without an NLA. 

{There are several experimental advances in the field that make the implementation of the analysed system here feasible in the short term. An early demonstration of the QS operation using heralded single-photon sources based on parametric down-conversion and avalanche photodiodes, as single-photon detectors, has already provided a proof-of-principle for the main building block of the system. With current technology, one can use higher quality single-photon sources based on quantum dot structures, and nanowire superconducting detectors for highly efficient low-noise photodetetion  \cite{Senellart_qdots:NatPhot2017,Cahall_SPDetectors:2017}. A combination of these two could bring down the internal noise in a QS module below a critical level that one can observe the benefits of deploying QSs in long-distance CV-QKD systems, as we have predicted in this work. This will be experimentally tested as part of our future work.} 

{The research conducted here can be further extended in several directions. Our study would, in particular, be highly relevant to analysing the performance of recently proposed  continuous-variable quantum repeater systems in \cite{Dias_Ralph_CVQRs}, which rely on a similar building block as we studied in this work. In their proposal, dual homodyne detection modules are used to connect different  blocks in the system. Considering the sensitivity to the excess noise in each leg of the system, it would be interesting to find out the regimes of operation in which a multi-hop CV repeater can be used for QKD purposes. One can compare the obtained key rates in this case with the already known benchmarks for the repeaterless links, i.e., the PLOB bound \cite{Pirandola_PLOB17}, as well as multi-node repeater setups \cite{Pirandola_RepeaterCapacity2019}. Another possible avenue for future work is to find better NLA schemes than QSs that better match the discrete modulation scheme used in this work. In fact, an alternative to QSs is a quantum comparison amplifier, which works on the basis of comparing the input coherent state with a known coherent state  \cite{Eleftheriadou_QAmp,Donaldson_ImpNLA_PRL}. Such an amplifier is still non-deterministic, but, it does not need  single-photon sources. Because a comparison amplifier can only amplify states that are chosen from a pre-known finite set of coherent states, it can possibly be a good fit to the QPSK-modulation protocol, where the number of transmitted coherent states is finite. Finally, one can also explore the use of numerical techniques \cite{PRXGhorai,Norbert_DM_CV} for key rate analysis, which can possibly better address the case of non-Gaussian attacks, and/or when analytical solutions become too cumbersome.}

\begin{acknowledgments}
The authors acknowledge partial support from the White Rose Research Studentship and the UK EPSRC Grant No. EP/M013472/1. S.P. would like to acknowledge funding from the European Union’s Horizon 2020 research and innovation program under grant agreement No 820466 (Continuous Variable Quantum Communications, `CiViQ'). All data generated in this paper can be reproduced by the provided methodology and equations.
\end{acknowledgments}

\appendix 

\section{Parameters of the co-variance matrix}
\label{app:CM-parameters}
In this section we calculate the triplet that quantifies the CM of our QS system, given in Eq.~\eqref{CM-TMSV}.

\subsection{Variance at Alice's side ($V_x$)}
By definition, and using the bipartite state in Eq.~\eqref{bipartite-state}, we have:
\begin{align}
V_x & = \mathrm{tr}(\widehat{\rho}_{03} \widehat{x}_0^2) = \frac{1}{4P^{\mathsf{PS}}}  \sum_{k=0}^{3} \sum_{l=0}^{3}  G_{kl} H_{kl} , 
\end{align}
where $\widehat x_0 = \widehat a_0 + \widehat a_0^\dag $ in Fig.~\ref{fig:EB-CV-QKD}, $ G_{kl} := \mathrm{tr}(|\psi_k\rangle_0\langle \psi_l |   \widehat{x}_0^2)  $ and $ H_{kl} := \mathrm{tr}(\widehat{\Omega}_3^{kl}) = \zeta_A^{kl}(0) $. We then find that:
\begin{align}
\label{H_func}
H_{kl} & = \zeta_A^{kl}(0) = a_{kl} e^{-\frac{T\alpha_k \alpha_l^\ast}{2F+1}} - \frac{1-\mu}{2F} e^{-\frac{T\alpha_k \alpha_l^\ast}{2F}}  \nonumber \\
a_{kl} & = \frac{2}{(2F+1)^3} \Big( (2F+1)^2 -\mu (2F+1)+ \mu T\alpha_k \alpha_l^\ast  \Big) .
\end{align}
One can then use the set of identities in Eq.~\eqref{state_identities} to work out the following expression:
\begin{align}
V_x = & 1+ \frac{\alpha^2}{\zeta_A^{00}(0)} \Big( \nonumber \\
& \delta_1 \big[ -A\sinh(\frac{T\alpha^2}{2F+1}) + B\cosh(\frac{T\alpha^2}{2F+1}) + C \sinh(\frac{T\alpha^2}{2F}) \big]   \nonumber \\
+  & \delta_2 \big[ A\cosh(\frac{T\alpha^2}{2F+1}) - B\sinh(\frac{T\alpha^2}{2F+1}) - C \cosh(\frac{T\alpha^2}{2F})  \big]   \nonumber \\
 + & \delta_3 \big[ -A\sin(\frac{T\alpha^2}{2F+1}) + B\cos(\frac{T\alpha^2}{2F+1}) + C \sin(\frac{T\alpha^2}{2F})  \big] /2 \nonumber \\
 - & \delta_4 \big[ A\cos(\frac{T\alpha^2}{2F+1}) + B\sin(\frac{T\alpha^2}{2F+1}) - C \cos(\frac{T\alpha^2}{2F}) \big]  /2 
\Big),
\end{align}
where 
$A = \frac{2}{(2F+1)^3} \Big( (2F+1)^2 -\mu (2F+1) \Big)$, 
$B = \frac{ 2 \mu T\alpha^2  }{(2F+1)^3}$,
$C = \frac{1-\mu}{2F}$,
$\delta_1 = \frac{\lambda_0}{\lambda_1}  + \frac{\lambda_2}{\lambda_3}$,    
$\delta_2  =  \frac{\lambda_1}{\lambda_2}  + \frac{\lambda_3}{\lambda_0}$,  
$\delta_3  =  \frac{\lambda_0}{\lambda_1}  - \frac{\lambda_2}{\lambda_3}$,     and
$\delta_4  =  \frac{\lambda_1}{\lambda_2}  - \frac{\lambda_3}{\lambda_0}$.
Note that for $\alpha=0$, $V_x=1$ is obtained. 

\subsection{Variance at Bob's side ($V_y$)}
The variance at the receiver's side can be computed as follows: 
\begin{align}
V_y & = \mathrm{tr}(\widehat{\rho}_{03} \widehat{x}_3^2) = \frac{1}{4P^{\mathsf{PS}}}  \sum_{k=0}^{3} L_{kk} ,
\end{align}
where, assuming $\xi_3=z+it$,
\begin{align}
L_{kk}  = & \mathrm{tr}(\widehat{\Omega}_3^{kk} \widehat{x}_3^2) \nonumber \\
= & - \zeta_A^{kk}(0,0) - \frac{d^2}{dt^2} \zeta_A^{kk}(0,t) \Big|_{t=0}  \nonumber \\
\frac{d^2}{dt^2} \zeta_A^{kk}(0,t) \Big|_{t=0} = & - b_k  e^{-\frac{T|\alpha_k|^2}{2F+1}}  + \frac{2(1-\mu)}{F} e^{-\frac{T|\alpha_k|^2}{2F}}   , 
\end{align}
with $\widehat x_3 = \widehat b_3 + \widehat b_3^\dag $ in Fig.~\ref{fig:EB-CV-QKD} and $b_k = \frac{8}{(2F+1)^3} \big( (2F+1)^2 -\mu (2F^2+3F+1) +\mu T |\alpha_k|^2 \big)$; hence,
\begin{align}
V_y = & \frac{L_{00}}{\zeta_A^{00}(0)} \nonumber \\
= &  \frac{1}{\zeta_A^{00}(0)}  \Big(  b_k  e^{-\frac{T|\alpha_k|^2}{2F+1}}  - \frac{2(1-\mu)}{F} e^{-\frac{T|\alpha_k|^2}{2F}}  \Big) -1 .
\end{align}
Note that for $\alpha=0$, $V_y=1$ is obtained. 

\subsection{Co-variance between Alice and Bob ($V_{xy}$)}
By definition, the co-variance between Alice and Bob is given by: 
\begin{align}
V_{xy} & = \mathrm{tr}(\widehat{\rho}_{03} \widehat{x}_0 \widehat{x}_3) = \frac{1}{4P^{\mathsf{PS}}}  \sum_{k=0}^{3} \sum_{l=0}^{3} N_{kl} S_{kl} , 
\end{align}
where $ N_{kl} := \mathrm{tr}(|\psi_k\rangle_0\langle \psi_l | \widehat{x}_0)$ is given in Eq.~\eqref{state_identities} and 
\begin{align}
S_{kl} = & \mathrm{tr}(\widehat{\Omega}_3^{kl} \widehat{x}_3)  \nonumber \\
= & -i  \frac{d}{dt} \zeta_A^{kl}(0,t) \Big|_{t=0}  \nonumber \\
=  & \frac{2 \sqrt{\mu(1-\mu)T} (\alpha_k+\alpha_l^\ast)}{(2F+1)^2} e^{-\frac{T\alpha_k \alpha_l^\ast}{2F+1}}  
\end{align}
One can then conclude that: 
\begin{align}
V_{xy} = & \frac{2\sqrt{\mu(1-\mu)T}\alpha^2}{P^{\mathsf{PS}}  (2F+1)^2} \big( \omega_1 \cosh(\frac{T\alpha^2}{2F+1}) \nonumber \\
&  - \omega_2 \sinh(\frac{T\alpha^2}{2F+1})  +  \omega_3 \cos(\frac{T\alpha^2}{2F+1})    \nonumber \\
&  -  \omega_4 \sin(\frac{T\alpha^2}{2F+1})  \big) ,
\end{align}
where 
$\omega_1 = \sqrt{\frac{\lambda_0}{\lambda_1}} + \sqrt{\frac{\lambda_2}{\lambda_3}}$,   
$\omega_2 = \sqrt{\frac{\lambda_1}{\lambda_2}} + \sqrt{\frac{\lambda_3}{\lambda_0}}$, 
$\omega_3 = \sqrt{\frac{\lambda_0}{\lambda_1}} - \sqrt{\frac{\lambda_2}{\lambda_3}}$,   and 
$\omega_4 = \sqrt{\frac{\lambda_1}{\lambda_2}} - \sqrt{\frac{\lambda_3}{\lambda_0}}$.
 It is seen that for $\alpha=0$, $V_{xy}=0$ is obtained. 

In the calculations of $G_{kl}$ and $N_{kl}$ we made use of the following identities:
\begin{align}
\label{state_identities}
|\psi_0\rangle = & \frac{1}{2} \big[ |\phi_0 \rangle +  e^{i\pi/4} |\phi_1 \rangle +   e^{i\pi/2} |\phi_2 \rangle + e^{3i\pi/4} |\phi_3 \rangle  \big],  \nonumber \\
\widehat{a}|\psi_0\rangle = & \frac{\alpha}{2} \big[ e^{i\pi/4} \sqrt{\frac{\lambda_0}{\lambda_1}} |\phi_0 \rangle +   e^{i\pi/2} \sqrt{\frac{\lambda_1}{\lambda_2}} |\phi_1 \rangle \nonumber \\
& + e^{i3\pi/4} \sqrt{\frac{\lambda_2}{\lambda_3}} |\phi_2 \rangle  - \sqrt{\frac{\lambda_3}{\lambda_0}} |\phi_3 \rangle \big] , \nonumber \\
\widehat{a}^2|\psi_0\rangle = & \frac{\alpha^2}{2} \big[ e^{i\pi/2} \sqrt{\frac{\lambda_0}{\lambda_2}} |\phi_0 \rangle +   e^{i3\pi/4} \sqrt{\frac{\lambda_1}{\lambda_3}} |\phi_1 \rangle \nonumber \\
& - \sqrt{\frac{\lambda_2}{\lambda_0}} |\phi_2 \rangle  - e^{i\pi/4} \sqrt{\frac{\lambda_3}{\lambda_1}} |\phi_3 \rangle \big] ,  \nonumber \\
|\psi_1\rangle = & \frac{1}{2} \big[ |\phi_0 \rangle +  e^{i3\pi/4} |\phi_1 \rangle +   e^{i3\pi/2} |\phi_2 \rangle + e^{i\pi/4} |\phi_3 \rangle  \big] , \nonumber \\
\widehat{a}|\psi_1\rangle = & \frac{\alpha}{2} \big[ e^{i3\pi/4} \sqrt{\frac{\lambda_0}{\lambda_1}} |\phi_0 \rangle +   e^{i3\pi/2} \sqrt{\frac{\lambda_1}{\lambda_2}} |\phi_1 \rangle \nonumber \\
& + e^{i\pi/4} \sqrt{\frac{\lambda_2}{\lambda_3}} |\phi_2 \rangle  - \sqrt{\frac{\lambda_3}{\lambda_0}} |\phi_3 \rangle \big] , \nonumber \\
\widehat{a}^2|\psi_1\rangle = & \frac{\alpha^2}{2} \big[ e^{i3\pi/2} \sqrt{\frac{\lambda_0}{\lambda_2}} |\phi_0 \rangle +   e^{i\pi/4} \sqrt{\frac{\lambda_1}{\lambda_3}} |\phi_1 \rangle \nonumber \\
& - \sqrt{\frac{\lambda_2}{\lambda_0}} |\phi_2 \rangle  - e^{i3\pi/4} \sqrt{\frac{\lambda_3}{\lambda_1}} |\phi_3 \rangle \big],  \nonumber \\
|\psi_2\rangle = & \frac{1}{2} \big[ |\phi_0 \rangle +  e^{-i3\pi/4} |\phi_1 \rangle +   e^{i\pi/2} |\phi_2 \rangle + e^{-i\pi/4} |\phi_3 \rangle  \big] , \nonumber \\
\widehat{a}|\psi_2\rangle = & \frac{\alpha}{2} \big[ e^{-i3\pi/4} \sqrt{\frac{\lambda_0}{\lambda_1}} |\phi_0 \rangle +   e^{i\pi/2} \sqrt{\frac{\lambda_1}{\lambda_2}} |\phi_1 \rangle \nonumber \\
& + e^{i\pi/4} \sqrt{\frac{\lambda_2}{\lambda_3}} |\phi_2 \rangle  - \sqrt{\frac{\lambda_3}{\lambda_0}} |\phi_3 \rangle \big] , \nonumber \\
\widehat{a}^2|\psi_2\rangle = & \frac{\alpha^2}{2} \big[ e^{i\pi/2} \sqrt{\frac{\lambda_0}{\lambda_2}} |\phi_0 \rangle +   e^{-i\pi/4} \sqrt{\frac{\lambda_1}{\lambda_3}} |\phi_1 \rangle \nonumber \\
& - \sqrt{\frac{\lambda_2}{\lambda_0}} |\phi_2 \rangle  - e^{-i3\pi/4} \sqrt{\frac{\lambda_3}{\lambda_1}} |\phi_3 \rangle \big],   \nonumber \\
|\psi_3\rangle = & \frac{1}{2} \big[ |\phi_0 \rangle +  e^{-i\pi/4} |\phi_1 \rangle +   e^{i3\pi/2} |\phi_2 \rangle + e^{-3i\pi/4} |\phi_3 \rangle  \big] , \nonumber \\
\widehat{a}|\psi_3\rangle = & \frac{\alpha}{2} \big[ e^{-i\pi/4} \sqrt{\frac{\lambda_0}{\lambda_1}} |\phi_0 \rangle +   e^{i3\pi/2} \sqrt{\frac{\lambda_1}{\lambda_2}} |\phi_1 \rangle \nonumber \\
& + e^{-i3\pi/4} \sqrt{\frac{\lambda_2}{\lambda_3}} |\phi_2 \rangle  - \sqrt{\frac{\lambda_3}{\lambda_0}} |\phi_3 \rangle \big],  \nonumber \\
\widehat{a}^2|\psi_3\rangle = & \frac{\alpha^2}{2} \big[ e^{i3\pi/2} \sqrt{\frac{\lambda_0}{\lambda_2}} |\phi_0 \rangle +   e^{-i3\pi/4} \sqrt{\frac{\lambda_1}{\lambda_3}} |\phi_1 \rangle \nonumber \\
& - \sqrt{\frac{\lambda_2}{\lambda_0}} |\phi_2 \rangle  - e^{-i\pi/4} \sqrt{\frac{\lambda_3}{\lambda_1}} |\phi_3 \rangle \big].  
\end{align}
 
\section{Calculation of Holevo Information}
 \label{app:HolevoInfo}
 For a CM in the following standard symplectic form 
\begin{align}
V_{AB}=
\left(\begin{array}{cc}
V_x \mathbbm{1} & V_{xy} \sigma_{\mathsf{z}} \\
V_{xy} \sigma_{\mathsf{z}} &  V_y \mathbbm{1}
\end{array}\right),  
\end{align}
 the Holevo information is upper bounded by:
 \begin{align}
 \label{app:Eq-Holevo}
 \chi_{EB}= g(\Lambda_1) + g(\Lambda_2)  -  g(\Lambda_3), 
 \end{align}
 where 
 $g(x)= (\frac{x+1}{2}) \log_2 (\frac{x+1}{2})  - \frac{x-1}{2}  \log_2 \frac{x-1}{2}  $
 and
 $ \Lambda_{1/2}=  \sqrt{ \big( W \pm \sqrt{W^2 - 4 D^2}\big)/2 }$ and $\Lambda_{3} = \sqrt{V_x D/V_y} $, with $W= V_x^2 + V_y^2 - 2 V_{xy}^2$ and $D= V_xV_y- V_{xy}^2$. Note that one can also take into account imperfect effects of the homodyne receiver. We however assume an ideal homodyne detection in this work.

\bibliography{biblio_dmCVQKD}
\bibliographystyle{apsrev4-1}

\end{document}